\documentstyle[epsf]{mn}

\newcommand{\dg} {^{\circ}}

\title[Timing analysis of the isolated neutron star RX
J0720.4-3125]{Timing analysis of the isolated neutron star RX
J0720.4-3125\thanks{Based on observations
obtained with {\sl XMM-Newton}, an ESA science mission with
instruments and contributions directly funded by ESA Member States and the
USA (NASA).}}

\author[]{Silvia Zane$^{1}$, Frank Haberl$^{2}$, Mark Cropper$^{1}$, 
Vyacheslav E. Zavlin$^{2}$, David Lumb$^{3}$,\and Steve Sembay$^{4}$ and
Christian
Motch$^{5}$\\
$^{1}$Mullard Space Science Lab, University College London,
Holmbury St. Mary, Dorking, Surrey, RH5 6NT, UK\\
$^{2}$ Max Planck Institut f\"ur Extraterrestrische Physik,
Giessenbachstrasse, D-85748 Garching, Germany\\
$^{3}$ Space Science Department, ESTEC, Posbus 299, Keplerlaan 1, Noordwijk
2200 AG, The Netherlands\\
$^{4}$ X-ray Astronomy Group, Department of Physics and Astronomy, University
of Leicester, Leicester LE1 7RH, UK \\
$^{5}$ Observatoire Astronomique, CNRS UMR 7550, 11 Rue de l'Universit\'e,
F-67000 Strasbourg, France}

\date{Received: }

\begin{document}

\outer\def\gtae {$\buildrel {\lower3pt\hbox{$>$}} \over
{\lower2pt\hbox{$\sim$}} $}
\outer\def\ltae {$\buildrel {\lower3pt\hbox{$<$}} \over
{\lower2pt\hbox{$\sim$}} $}
\newcommand{\ergscm} {ergs s$^{-1}$ cm$^{-2}$}
\newcommand{\ergss} {ergs s$^{-1}$}
\newcommand{\ergsd} {ergs s$^{-1}$ $d^{2}_{100}$}
\newcommand{\pcmsq} {cm$^{-2}$}
\newcommand{\ros} {\sl ROSAT}
\newcommand{\xmm} {\sl XMM-Newton}
\newcommand{\exo} {\sl EXOSAT}
\def\rchi{{${\chi}_{\nu}^{2}$}}
\newcommand{\Msun} {$M_{\odot}$}
\newcommand{\Mwd} {$M_{wd}$}
\def\Mdot{\hbox{$\dot M$}}
\def\mdot{\hbox{$\dot m$}}
\def\mincir{\raise -2.truept\hbox{\rlap{\hbox{$\sim$}}\raise5.truept
\hbox{$<$}\ }}
\def\magcir{\raise -4.truept\hbox{\rlap{\hbox{$\sim$}}\raise5.truept
\hbox{$>$}\ }}

\maketitle

\begin{abstract}

We present a combined analysis of {\sl XMM-Newton}, {\sl Chandra} and {\sl
Rosat} observations of the isolated neutron star RXJ0720.4-3125, spanning
a total period of $\sim 7$ years. We develop a maximum likelihood
periodogramme for our analysis based on the $\Delta
C$-statistic and the maximum likelihood method, which are appropriate 
for the treatment of sparse event lists. Our results have been checked
{\it a posteriori} by folding a further {\sl BeppoSAX}-dataset with the
period predicted at the time of that observation: the phase is found to be 
consistent. 

The study of the spin history and the measure of the spin-down rate is of
extreme importance in discriminating between the possible mechanisms
suggested for the nature of the X-ray emission. The value of $\dot P$,
here measured for the first time, is $\approx 10^{-14}$~s/s. This value 
can not be explained in terms of torque from a fossil disk. When
interpreted in terms of dipolar losses, it gives a
magnetic field of $B \approx 10^{13}$~G, making also implausible that the
source is accreting from the underdense surroundings.
On the other hand, we also find unlikely that the field decayed 
from a much larger value ($B\approx 10^{15}$~G, as expected for a
magnetar powered by dissipation of a superstrong field) since this
scenario predicts a source age of $\approx 10^4$~yrs, too young to match 
the observed X-ray luminosity. The observed properties are more
compatible with a scenario in which the source is $\approx
10^6$~yrs old, and its magnetic field has not changed substantially
over the lifetime. 

\end{abstract}

\section{Introduction}
\label{int}

RXJ0720.4-3125 is a nearby, isolated neutron star (NS) that was originally
discovered with ROSAT during a systematic survey of the galactic
plane (Haberl et al., 1997). The source exhibits all the common
characteristics
of the other six ROSAT NS candidates (hereafter dim NSs, see e.g. Treves
et al., 2000 for a review): a blackbody-like, soft spectrum with $kT\sim
80$~eV; an exceedingly
large X-ray to optical flux ratio; a low X-ray luminosity, $L_X\approx
10^{30}-10^{31} {\rm erg\,s}^{-1}$; a low column density and no evidence for a
binary companion.  An optical counterpart has been identified (Motch \&
Haberl, 1998; Kulkarni \& van Kerkwijk, 1998).

Pulsation associated with the spin period has been observed in four dim
NSs: RXJ0720.4-3125 ($P \approx 8.39$~s),
RXJ0420.0-5022 ($P \approx 22.69$~s), RXJ0806.4-4123 ($P \approx 11.37$~s)
and RBS1223 ($P \approx 5.16$~s). The pulse shape of RXJ0720.4-3125 is
almost sinusoidal (Haberl et al., 1997; Cropper et al., 2001).  

The true nature of the mechanism powering the X-ray emission from dim
NSs is still unclear. Until a few years ago, these sources were thought to
constitute a stand alone class and two mechanisms were proposed for their
emission: either accretion from the interstellar medium onto an old
neutron star or release of thermal radiation from a younger, cooling
object (see e.g. Pavlov et al., 1996; Treves et al., 2000; Motch, 2001 and
references therein). More recently, it
has been noted that the period of RXJ0720.4-3125 is somewhat unusual
and falls in the very narrow range in which the anomalous X-ray pulsars
(AXPs) periods cluster ($P\sim 6-12$ s, see e.g. Mereghetti, 2000). This
hints at a possible
evolutionary link between dim
NSs, AXPs, and soft gamma-ray repeaters (SGRs, see e.g. Heyl \& Kulkarni,
1998; Heyl \& Hernquist, 1998; Colpi et al., 2000; Alpar, 1999, 2000,
2001). Two kind of ``unified''
scenarios have been then proposed. In the first one, the common mechanism
powering the three classes of objects is dissipation of a decaying,
superstrong magnetic field ($B\magcir 10^{14}-10^{15}$~G). In this case
dim NSs represent the descendants of SGRs and AXPs, and RXJ0720.4-3125
may be one of the closest old magnetars.
Alternatively, all three classes may contain neutron stars with lower
(canonical) magnetic field ($B \approx
10^{12}~G$) endowed by a fossil disk (Alpar, Ankay \& Yazgan, 2001;
Alpar, 2001). In
this
interpretation, dim NSs would be in the propellor phase and would be the
progenitors of AXPs and SGRs, the latter having entered an accretion
phase.

\begin{table*}
\begin{center}
\begin{tabular}{llllrrl}
\hline
Date & Observatory & Instrument & Exposure    & Exposure & Effective & Label \\
     &             &         & identification & duration & exposure  &       \\
     &             &         &                & (s)    & (s)     &       \\
\hline
1993 Sep 27 & {\sl Rosat}      & PSPC & rp300338n00 &  11980 &  3221 & R93 \\
1996 Apr 25 & {\sl Rosat}      & HRI &  rh300508n00 &   7838 &  3566 & R96a \\
1996 May 7  & {\sl Rosat}      & HRI &  rh180100n00 &   7743 &  3125 & R96b \\
1996 Sep 27 & {\sl Rosat}      & HRI &  rh300508a01 &   1498 &  1409 & R96c \\
1996 Nov 3  & {\sl Rosat}      & HRI &  rh400884n00 &  65698 & 33569 & R96d \\
1998 Sep 27 & {\sl Rosat}      & HRI &  rh400944n00 & 460195 &  3566 & R98 \\
2000 Feb 1  & {\sl Chandra}    & HRC-S(LETG 1st order) & 348+349+745 & 305528 &
  37635 & Ch00 \\
2000 May 13 & {\sl XMM-Newton} & MOS1 + thin filter & 0124100101-001 & 61648 &
  61648 & X00a \\
            &       & MOS2 + thin filter & 0124100101-002 & 61648 & 61648 \\
            &       & PN + thin filter   & 0124100101-003 & 62425 & 62425 \\
2000 Nov 21 & {\sl XMM-Newton} & MOS1 + medium filter & 0132520301-007 & 17997
& 17997 & X00b \\
            &       & MOS2 + medium filter & 0132520301-008 & 17994 & 17994 \\
            &       & PN + medium filter &  0132520301-003 &25651 & 25651 \\
\hline
%            &       &                    &                &       &       \\
1997 Mar 16 & {\sl BeppoSAX}   & LECS & LECS\_20079001 & 99418 &   & S97    \\
\hline
\end{tabular}
\caption{The ROSAT, Chandra and {\sl XMM-Newton}
observations of RXJ0720.4-3125 used in this paper.}
\label{obslog}
\end{center}
\end{table*}

In any model involving magnetars, proton cyclotron features are expected
to lie in the X-ray range, while electron cyclotron lines are expected
in the same band for
canonical magnetic field.
According to Zane et al. (2001), for surface magnetic fields strengths of
$\sim 10^{14}-10^{15}$~G spectra exhibit a distinctive absorption feature
at the proton cyclotron energy $\sim 0.63 (B/10^{14}\, {\rm G})$
keV. The required resolving power is $\approx
100$, therefore the detection of this feature is well within the
capabilities of {\sl XMM-Newton} grating spectrometers.
Recently, Paerels et al.~(2001) have presented spectra of
RXJ0720.4-3125 using {\sl XMM-Newton}, and found that
there is no evidence for absorption lines and edges in the X-ray spectrum.
Unless different atmospheric effects may concur in suppressing the
feature, when taken straightforward the absence of electron or proton
cyclotron resonances in the RGS band excludes a
range of average magnetic field strengths, $B \approx (0.3-2)\times
10^{11}$~G and $(0.5-2)\times 10^{14}$~G.

Based on the same {\sl XMM-Newton} observation, Cropper et al.~(2001) 
presented the pulse-shape analysis of RXJ0720.4-3125,
modelling spin pulse intensity and hardness ratio profiles.
By assuming that the source of the flux variation is the changing
visibility of the heated magnetic polar caps, they derived an upper limit
on the cap size and showed that a polar cap larger than $\sim 60\dg
-65\dg$ can be rejected at a confidence level of 90\%. Whatever the
mechanism, the X-ray emitting region is therefore confined to a
relatively small fraction of the star surface. It is worth noticing that
such a small hot region can not be explained only in terms of 
non-uniformity in the distribution of the surface cooling temperature
induced by an high magnetic field (Greenstein \& Hartke~1983; Page~1995).
The  temperature gradient which is induced by the $B$-dependence of
the thermal conducivity of the neutron star crust is relatively
smooth and,  even in the (unrealistic) limit case $B\to \infty$ the
associated blackbody luminosity drops only by a factor 1/2 at $\sim
60 \dg$ and by an order of magnitude at $\sim 77 \dg$. By performing pulse
phase spectroscopy, Cropper et al.~(2001) also found that the observed
hardness ratio is softest slightly before flux maximum. The same
characteristic was later discovered by Perna et al. (2001) in the
spectra of AXPs. 
Cropper et al. ~(2001) suggested two possible
explanations for this effect: their best-fitting model is based on
radiation beaming, while an alternative one assumes a spatially variable
absorbing matter, co-rotating in the magnetosphere. The latter may be
indeed the case if the star is propelling matter outward (Alpar,
Ankay \& Yazgan~2001).

Further information about the nature of this puzzling source can be
obtained by the spin history. In magnetars magneto-dipolar radiation will
cause rapid spin-down at a rate $\dot P \approx 10^{-11}(B/10^{14}\, {\rm
G})^2/P$ ss$^{-1}$, and it has been the positive detection of a secular
spin-down of this order that has suggested the association of AXPs and
SRGs with ultra-magnetized NSs (Kouveliotou et al., 1998, 1999; Thompson
et al., 2001). The preliminary measure of $\dot P$
published by Haberl et al. (1997) for RXJ0720.4-3125 is uncertain to a
considerably large value, and does not allow spin-up and spin-down
to be discriminated. Very recently, Hambaryan et al. (2002) presented the
{\it first evidence of large spin-down rate} in a dim
NS, RBS1223. Their measure is partially based on ROSAT data where the
detection of the period was not highly significant, and also the value
($\dot P = 1.35 _{-0.67}^{+0.69}
\times 10^{-11}$~s/s) is
still compatible with being due to torque from a fossil disk. However, it
is certainly worth
noticing that, when taken face value and
interpreted in terms of dipolar rotational losses, this implies the
presence of a superstrong magnetic field, $B \approx
(3.5-6.5) \times 10^{14}$~G.

A similar measure of $\dot P$ for the closest pulsating
candidate, RXJ0720.4-3125, is therefore
of extreme importance, as well as the accurate tracking of its spin
history.

This is the first opportunity to investigate the pulse timing in the {\sl
XMM-Newton} data, since at the time of the previous papers (Paerels et al. 
2001,
Cropper et al., 2001) the timing correlation files were not sufficiently well
determined. Here we present a combined analysis of {\sl XMM-Newton}, {\sl
Chandra} and {\sl Rosat} data, spanning a period of $\sim 7$ years.

\section{Observations}
\label{obs}

The log of observations used for this study is given in Table~\ref{obslog}.

RXJ0720.4-3125 has been observed twice by {\sl XMM-Newton}: first on 2000 May
13, during the Calibration/Performance Verification phase (Paerels et al.
2001,
Cropper et al. 2001) and later on 21 November 2001, again for calibrations. For
the analysis presented in this paper we have used data from all three EPIC
cameras on both epochs. The data were reduced using the {\sl XMM-Newton}
Scientific Analysis System (SAS) version 5.1. Source counts in the range
0.12 to 1.2~keV (PN) and  0.1 to 2.0~keV (MOS) were extracted in an
aperture of 40 arcsec for the PN X00a, 30
arcsec for PN X00b, and 30 arcsec for the
MOS cameras. Events with patterns 0-4 (single and double events) were
selected for MOS, while patterns 0-12 (all the valid PN patterns) were
retained for PN. Data arriving during episodes of high background which
were experienced in the 2000 May 13 observation were excluded. Also in
this observation, some PN event times were noted not to be
integer multiple of the frame times, but delayed by 1 second: such events
(in total about 34~ks from the 50~ks) 
were corrected in the event list. Finally, times were converted to 
Barycentric Dynamical Time (TDB) at solar system barycentre.

The {\sl Chandra} observations were made over three days starting on 2000 Feb 1
using the HRC-S with LETG. The events were extracted from zero-order
images of the source (namely, from circles of 2 arcsec radius 
centered at the star's position). The dispersed
data (1$^{st}$ and higher orders) are strongly contaminated
by background (about 50-60\%) and are not useful for this
task. On the other hand, the background contamination in the zero-order
data is negligible, about 2-3\%. Also, the zero-order data
are not piled-up: this effect can appear only at larger count rates
($\sim 10^5$~ct/s).     

The {\sl Rosat} observations were taken over a period of five years in 1993,
1996 and 1998. An earlier report of the analysis of the 1993 and 1996 data is
in Haberl et al. (1997). The data were re-extracted and re-analysed using
the
EXSAS software system to make use of the most recent knowledge of spacecraft
clock corrections. The 1998 data were taken shortly before a spacecraft
emergency safe-hold, and do not benefit from a clock calibration: timing
corrections are therefore extrapolated from the last available
calibration. Timings may therefore be incorrect up to several seconds
(W. Becker, private communication): this
means that the reference of this observation to the rest of the dataset is
problematic. However, no significant drifts are expected over the duration of
the observation itself, so that the data are still useful in a standalone
analysis.

Although the countrate is too low for our main analysis, we have also extracted
the SAX LECS observation of RX J0720.4--3125 from the SAX data archive (the
source was not visible in the other SAX instruments) for use as an {\it a
posteriori} check on our derived timing parameters. Here we used the event
list
from the pipeline, correcting the times to the solar system barycentre using
the FTOOL earth2sun. This neglects the light travel time from Earth
centre to the satellite in low Earth orbit, $\sim 20$~ms, which is
negligible for the purposes of our {\it a postiori} check.

The major datasets in Table~\ref{obslog} are from the two {\sl XMM Newton}
observations which have high count-rates (particularly the EPIC-PN), and the
long 1996 Nov 3 {\sl Rosat} data. The 1998 {\sl Rosat} and the {\sl Chandra}
observations are valuable by nature of their several day durations.

\begin{figure}
\begin{center}
\setlength{\unitlength}{1cm}
\begin{picture}(20,5)
\put(-2.0,-2.4){\includegraphics{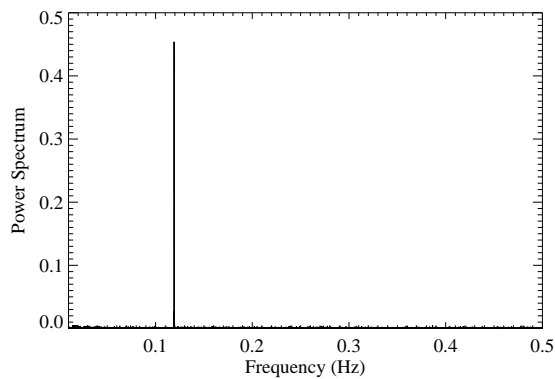}}
\end{picture}
\end{center}
\caption{The Rayleigh transform of the X00a PN dataset over the period
range 100~s to 2~s. This shows only the single frequency at 8.391~s 
(0.11917 Hz), as reported in Haberl et al. (1997).}
\label{Rayleigh}
\end{figure}

\section{Time Series Analysis}
\label{analysis}

\begin{figure*}
\begin{center}
\setlength{\unitlength}{1cm}
\begin{picture}(18,10)
\put(-2,-4){\includegraphics{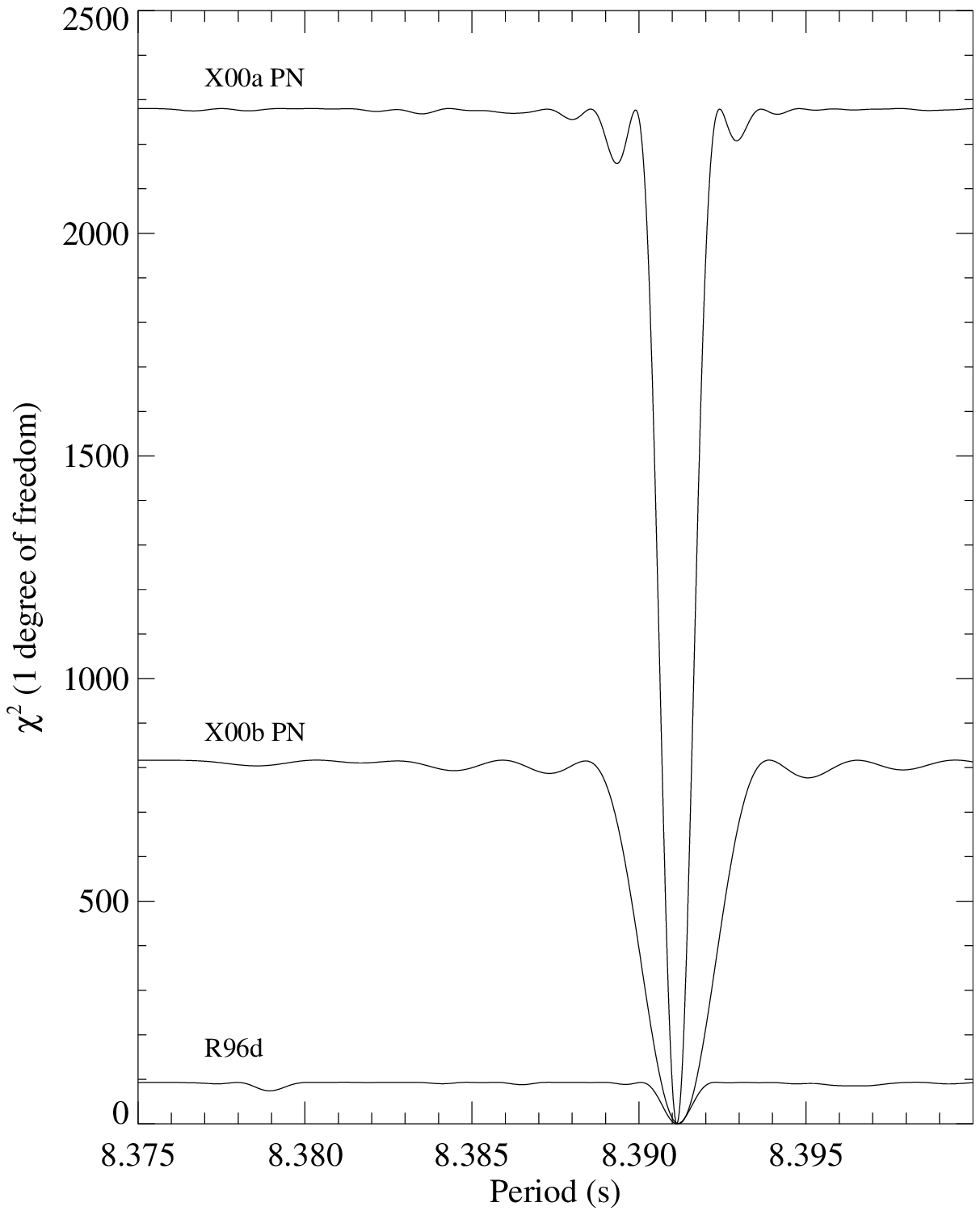}}
\put( 7,-4){\includegraphics{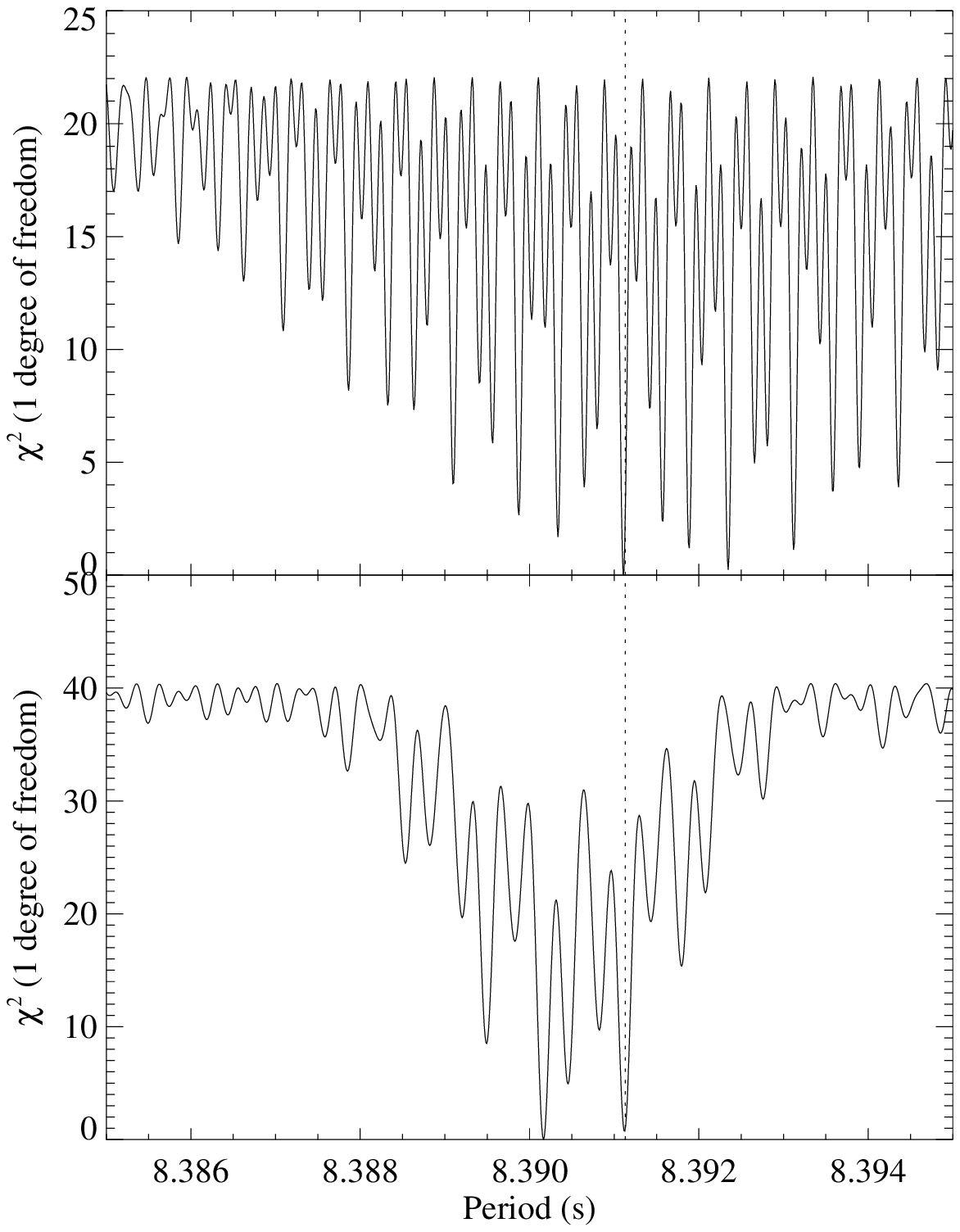}}
\end{picture}
\end{center}
\caption{(Left) Maximum likelihood periodogrammes (MLP) for three long
datasets, R96d, X00a (PN) and X00b (PN), showing the unambiguous detection of a
periodicity at 8.391~s. These constrain the selection of the strongest and
second-strongest dips in the MLPs for the R98 and Ch00 datasets respectively
(right). The vertical line denotes a period of 8.39113~s. The 68\% and
90\%
confidence levels are at $\chi ^2 = 1.0$ and 2.71 for one degree of freedom.}
\label{stage1}
\end{figure*}

Our data originate from instrumentation with widely differing sensitivities:
typical count rates vary from 1 count every $\sim 3$~s for {\sl Rosat}
HRI to
$\sim 6$ counts/s for {\sl XMM-Newton} PN. However, none of these count rates
is sufficiently high for a normal distribution of counts to be expected in a
time bin of adequate time resolution. The body of literature addressing
discrete Fourier Transforms (for example Deeming 1975, Scargle 1982,
Schwarzenberg-Czerny 1998) is not directly applicable for our analysis. The
treatment of sparse data and event list data is generally accomplished using
the Rayleigh Transform (for example de Jager 1991, Mardia 1972). It should be
noted that this transform is numerically the same as the discrete Fourier
Transform.

It will be important in our analysis to define confidence intervals to the
derived quantities, in particular the period. In the case where sufficient
counts per sample are available, this is generally achieved by least squares
fits to the data (for example Kurtz 1985), but this is not possible for
sparse
or event list data. In this case it is necessary to use the more general
maximum likelihood technique, which makes no assumptions on the data
distribution (Cash 1979, Lampton, Margon \& Bowyer 1976). Using this, Cash
(1979) defined the $C$- and $\Delta C$-statistics. The $\Delta C$-statistic
(Cash 1979) has the considerable advantage of being distributed as $\chi ^2$,
so that the required confidence intervals can be identified directly from the
plot.

The maximum likelihood technique depends on a model probability distribution,
which must be specified {\it a priori}. Haberl et al. (1997) and Cropper et
al. (2001) find the flux variation of RXJ0720.4-3125 to be almost sinusoidal,
so in our particular case we can take a sinusoidal variation superimposed on a
mean value as an appropriate model. The single frequency and absence of
harmonics in the Rayleigh transform of the X00a~PN dataset shown in
Figure \ref{Rayleigh} provides the justification explicitly for this
approach. The maximum likelihood is then computed
simply as equation (35) of Bai (1992). The normalisation of the model is
unimportant, as it results in a constant term in the C-statistic, which then is
eliminated in the $\Delta C$-statistic. In our case we have two free
parameters: normalised amplitude and phase. In general, when performing maximum
likelihood fits, the best fit free parameters have to be determined by whatever
method may be most appropriate. However, in the case of sinusoids, as noted by
Bai (1992), their orthogonality allows them to be determined directly from the
Rayleigh Transform (or Lomb-Scargle periodogramme).

The computation of the $\Delta C$-statistic is therefore only slightly more
complicated than the periodogramme. It starts in the same way, by a scan
through period, calculating the Rayleigh Transform to provide the amplitude and
phase at each period. Then at each period the maximum likelihood is calculated
using equation (35) of Bai (1992). Finally from equations (4) and (7) of
Cash
(1979),
$\Delta C$ is simply twice the difference between the maximum likelihood at a
particular period, and the maximum value over the whole frequency range. This
results in an inverted ``periodogramme'', with minimum value of zero. The
permitted period range is that between the confidence levels appropriate for
$\chi ^2$ with the appropriate number of degrees of freedom (1 or 2 depending
on whether $\dot P$ is included as a parameter -- see section
\ref{spin}). 

Details of the procedure are given in Appendix A.

With a period of 8.391~s (Haberl et al. 1997), more than $10^7$ cycles
have
elapsed over the time span of our data set. It is therefore essential to
incorporate a $\dot P$ term in our analysis techniques when combining datasets
at different epochs. The search then becomes a $(P_0 \, ,\dot P)$ search,
where $P_0$
is the period at the start of the first dataset (the 1993 {\sl Rosat}
observations), and the instantaneous period $P = P_0 + \dot P (t-t_0)$ is
computed at each point in the grid. Here $t_0$ is the time at the start of the
R93 run.

As a more general comment, we recommend this maximum likelihood periodogramme
(hereafter MLP) for time series analysis. The uncertainty in a period
determination is not evident directly from the width of a peak in a power
spectrum computed by classical discrete Fourier transform, Lomb-Scargle
periodogramme or Rayleigh transform, as this is set by the window function
({\it c.f.} Scargle 1982 appendix D).  However, it {\it is} directly available from
the MLP $\Delta C$ statistic: the inverted peak from a sinusoidal signal in
data with high signal-to-noise ratio will be narrower than the peak from low
signal-to-noise data and the $\chi^2$ can be read directly from the vertical
axis (see {\it e.g.} Figure \ref{stage1}). Furthermore, as noted above,
the MLP
technique requires an appropriate model, which may be unknown {\it a
priori}. In this case the signal can be reconstructed from the harmonic content
of the Rayleigh transform or Lomb-Scargle periodogramme, or from the phase
folded data, and an appropriate model generated. This model can then be used in
the calculation of the MLP, incorporating all of the harmonic information
explicitly.

\section{Measure of The Spin Period}
\label{spin}

The only measure of $P$ and $\dot P$ presented in the literature for
RXJ0720.4-3125 was given by Haberl et al. (1997) using {\sl Rosat}
data. The period was determined to be $8.39115\pm0.00002$ seconds from the 1996
Nov 3 dataset. Based on a linear fit of the data spanning 3 years, these
authors derived a 90 \% confidence range of $-6.0 \times 10^{-12}$ to $0.8
\times 10^{-12}$ for $\dot P$, compatible with no period change.

It is computationally unfeasible to explore this range in the $P_0 \dot P$ 
plane
in a single analysis of the entire dataset to determine refined values for
$P_0$ and $\dot P$. A reduction in the range is required. This can be
accomplished initially by extending the Haberl et al. (1997) analysis to
include
the {\sl XMM-Newton} data. This is sufficiently accurate to identify the
correct peak in the alias patterns in the extended duration R98 and Ch00 data,
leading to a further limiting of the permitted range in the $P \dot P$
plane. The next stage is a grid search in $P$ and $\dot P$ over this restricted
range for the 1993 and 1996 combined {\sl Rosat} data only. The final stage is
a grid search using all data in the restricted zone for which there are
acceptable solutions from this {\sl Rosat} analysis.

\subsection{Individual datasets}
\label{spin:indiv}

We begin by performing an MLP assuming $\dot P=0$ on each of the longer
pointing (R93, R96d, X00a, X00b) in the period range (8.375-8.400)~s,
incorporating the value of 8.39115~s found by Haberl et al. (1997). The
results are shown in Figure \ref{stage1} (left). As we can see, there is no
ambiguity in the period determinations from the R96d, X00a and X00b datasets;
the best fit frequencies and their uncertainties are reported in Table
\ref{periods}. A linear least square fit using the 68\% formal errors from the
MLP results in a value of $P_0 = 8.39113\pm0.00011$~s, $\dot
P=0.0\pm5.5\times10^{-13}$ s/s, where, as throughout, $P_0$ is referenced
to the start of the R93 run.

\begin{table}
\begin{center}
\begin{tabular}{llll}
\hline
Label & JD start & Period & 68\% Uncertainty \\
      &  (TDB)   &  (s) & (s)     \\
\hline
R93       & 2449257.684706 & 8.391201 & 0.000450       \\
R96a      & 2450199.141848 & --       &                \\
R96b      & 2450211.010212 & --       &                \\
R96c      & 2450354.488187 & --       &                \\
R96d      & 2450391.419553 & 8.391137 & 0.000063       \\
R98       & 2450923.523982 & 8.391109 & 0.000013       \\
Ch00      & 2451575.771501 & 8.391121 & 0.000020       \\
X00a PN   & 2451677.604110 & 8.391133 & 0.000021       \\
X00a MOS1 & 2451677.614469 & 8.391103 & 0.000043       \\
X00a MOS2 & 2451677.614474 & 8.391120 & 0.000040       \\
X00b PN   & 2451870.308664 & 8.391181 & 0.000050       \\
X00b MOS1 & 2451870.390808 & 8.391041 & 0.000167       \\
X00b MOS2 & 2451870.390823 & 8.391032 & 0.000181       \\
\hline
\end{tabular}
\caption{Best fit periods to the longer individual datasets, assuming
$\dot P=0$.}
\label{periods}
\end{center}
\end{table}

\begin{figure*}
\begin{center}
\setlength{\unitlength}{1cm}
%\fbox{
\begin{picture}(18,14)
\put(-2.0,-2.4){\includegraphics{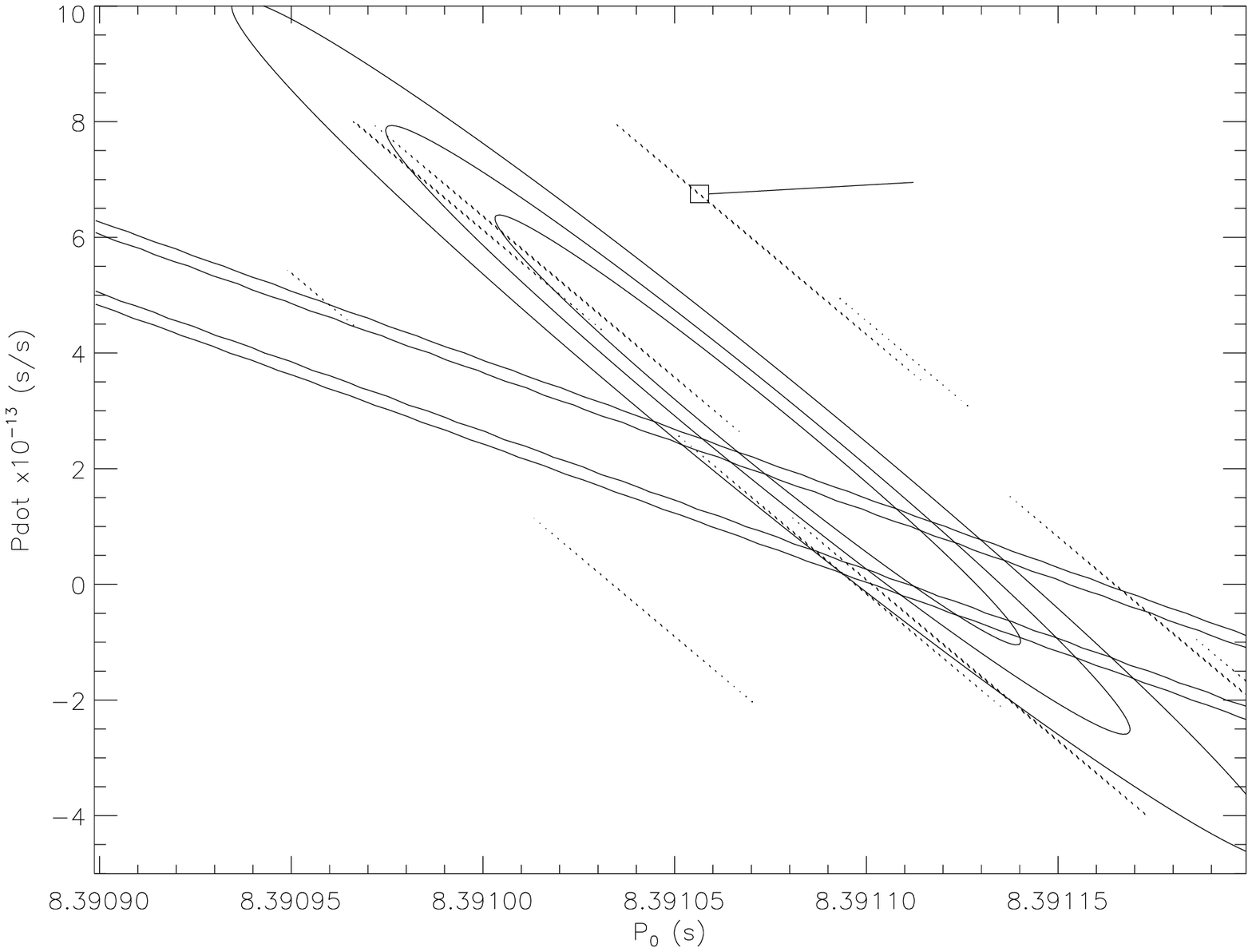}}
\put(11.65,7.15){\includegraphics{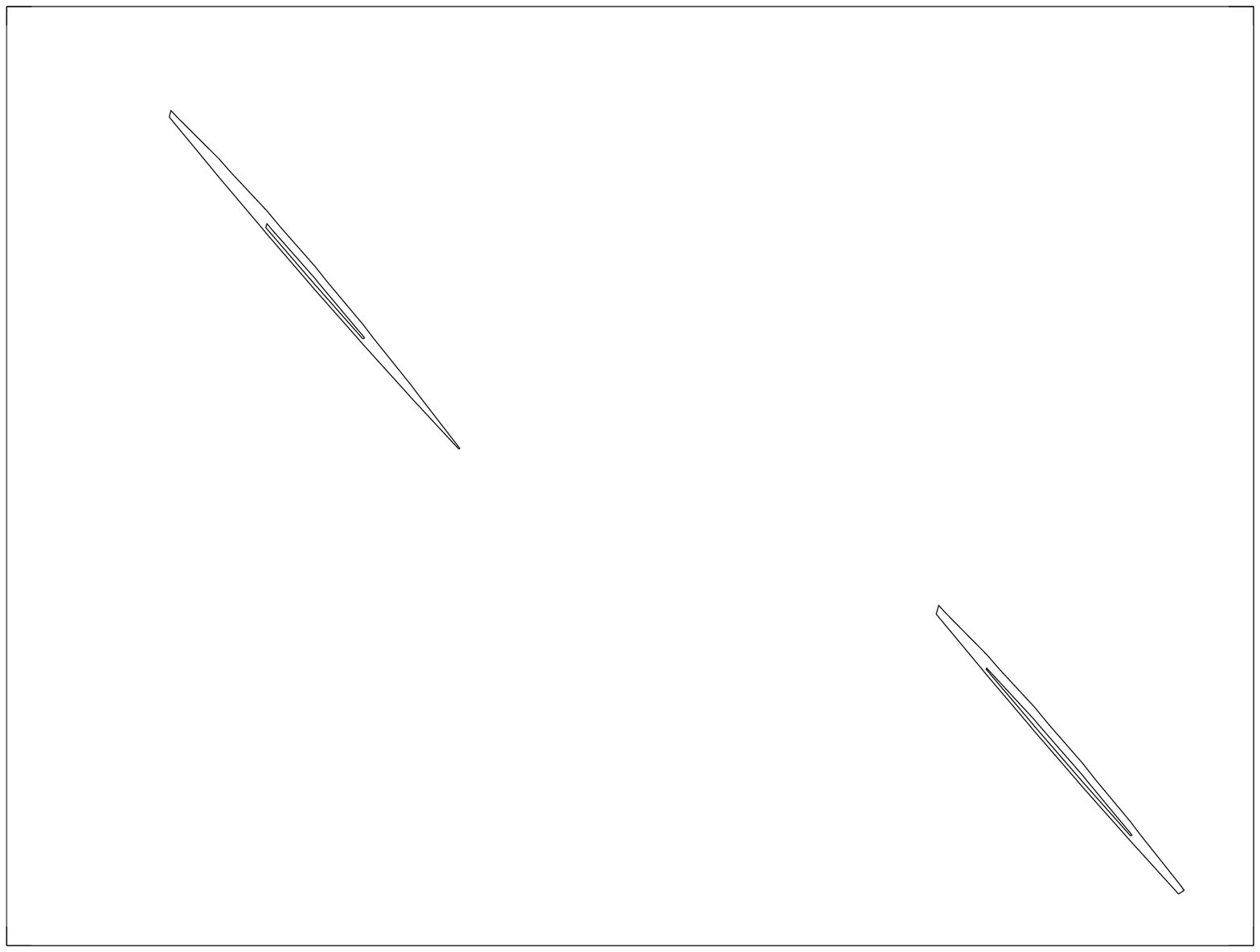}}
\end{picture}
%}
\end{center}
\caption{The 68\%, 90\% and 99\% contours for $P_{0}$ and $\dot P$ linear least
squares fit to the individual best fit periods in Table \ref{periods}
(continuous elliptical regions).  Overplotted on the same scale are the 68\%
and 90\% contours from the MLPs to the X00a PN (defined by continuous parallel
lines) and the combined 1993 and 1996 {\sl Rosat} datasets. The latter are
broken up into tiny elliptical regions corresponding to the 68\% and 90\%
confidence levels for the different aliases: see the enlargement inset in the
top right hand part of the figure.}
\label{stage2}
\end{figure*}

The proximity in time between the Ch00 and X00a observations, together with the
above upper limit in $\dot P$, permits an unambiguous determination of the
second highest peak in the Ch00 power spectrum (Figure
\ref{stage1}, right). The same is true for the R98 observations, in which
the highest peak
is selected. Adding these to the linear least squares fit results in a value of
$P_0 = 8.39107 \pm 0.00005$ and $\dot P =
2.7\times10^{-13}\pm2.5\times10^{-13}$. As may be expected, the $P_0$ and $\dot
P$ are highly correlated. The 68, 90 and 99\% confidence intervals are shown in
Figure \ref{stage2}.

The duration and high count-rate of the X00a PN run alone provides a strong
constraint in the $(P_0 \, ,\dot P)$ plane. The MLP 68\% and 90\%
confidence
intervals are also shown in Figure \ref{stage2}. This indicates that acceptable
values for $P_0$ and $\dot P$ lie within the range 8.39106 to 8.39115~s 
and $-1$ to $ +2 \times 10^{-13}$ s/s, respectively. We checked the effect
of a $10^{-7}$ error in the {\sl XMM-Newton} clock calibration, as
reported recently (Kuster, M., 2001, reported at ``New Visions of the
X-ray Universe in the {\sl XMM-Newton} and {\sl Chandra} Era''). The
change in the confidence region is negligible.  

\begin{figure*}
\begin{center}
\setlength{\unitlength}{1cm}
\begin{picture}(18,10)
\put(-1,-1.5){\includegraphics{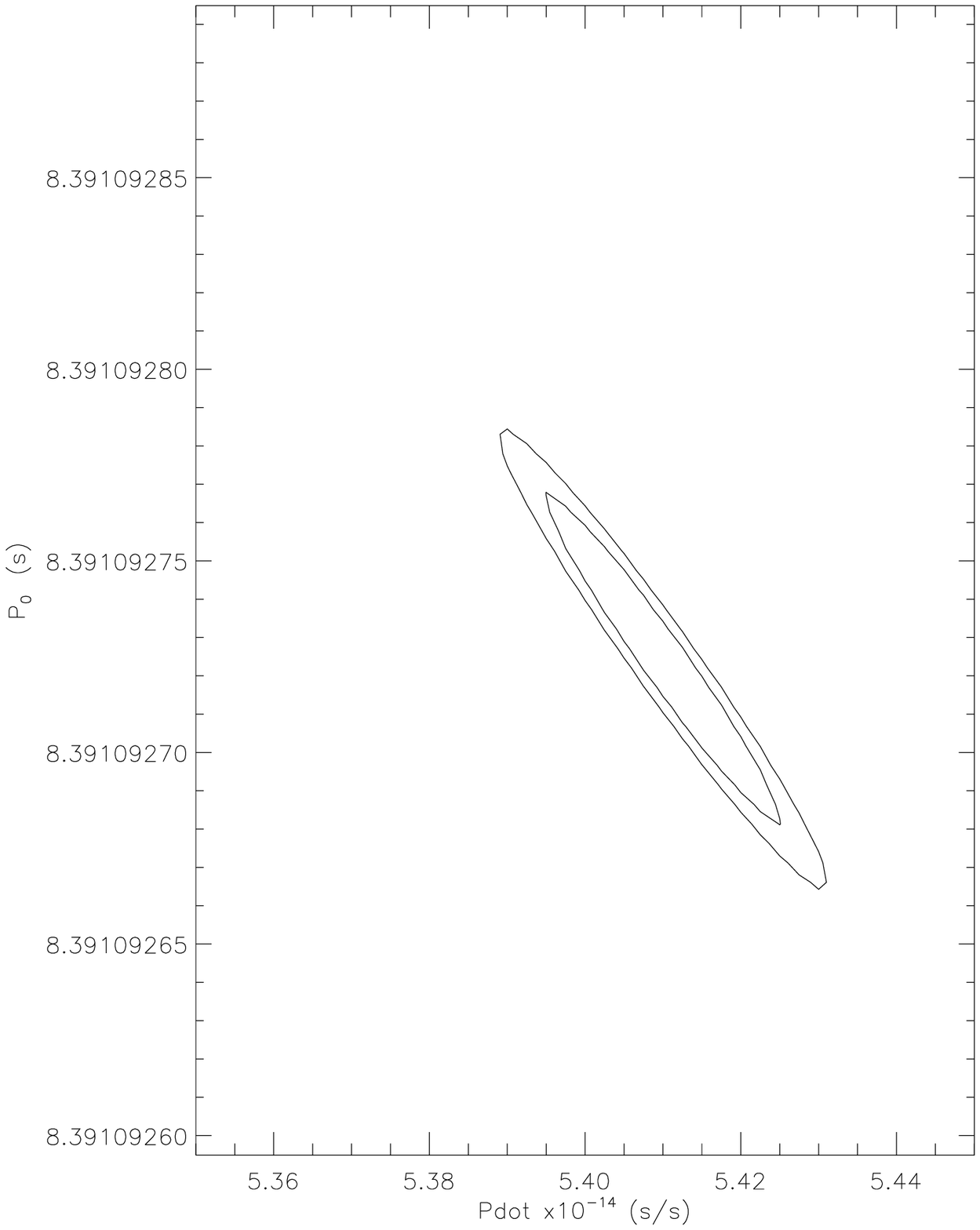}}
\put( 7,-1.5){\includegraphics{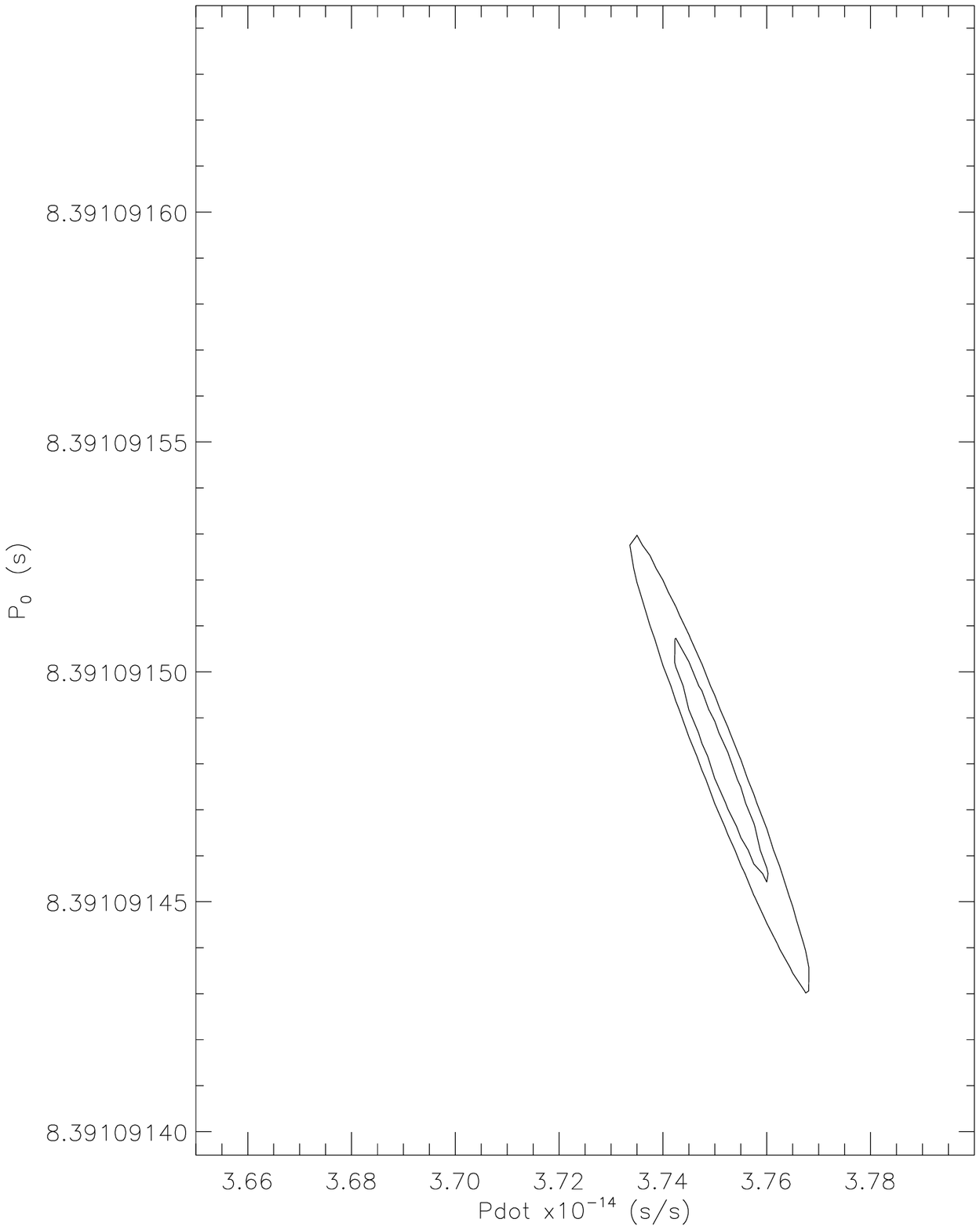}}
\put( 6.5,9){(1)}
\put(14.4,9){(2)}
\end{picture}
\end{center}
\caption{The 68\% and 90\% MLP contours for $P_{0}$ and $\dot P$ to the
complete dataset except for the R98 data, for the two acceptable  $(P_0 \,
,\dot P)$ values (1) and (2). Note that the contours in (2) are referenced
to the best
global fit derived for (1).}
\label{stage3}
\end{figure*}

\subsection{Combined {\sl Rosat} dataset}
\label{spin:rosat}

The $P_0$ $\dot P$ range identified in Figure \ref{stage2} is sufficiently
restricted for an MLP to be performed on the combined 1993 and 1996 {\sl Rosat}
datasets. The use of combined datasets requires coherent phasing to be
maintained over the whole dataset (coherent phasing was required only {\em
within} each of the individual datasets used until this point).  The
$(P_0 \, ,\dot P)$ plane was searched over between $8.39095 \leq P_0 \leq
8.39120$~s, and
$-4\times 10^{-13} \leq \dot P \leq 8 \times 10^{-13}$~s/s, exceeding the
parameter
range for the 90\% confidence limits from the linear least squares fit. The
resulting 68\% and 90\% confidence contours break up into small regions
(aliases) distributed along lines in the $(P_0 \, ,\dot P)$ plane. These
are also
shown in Figure \ref{stage2}

As is evident, there is overlap between the {\sl Rosat} 90\% confidence
interval contours and the 99\% contours of the linear least squares fit to the
individual data (but not quite at the 90\% level). This overlap region is also
consistent with the X00a PN confidence intervals.

\begin{figure*}
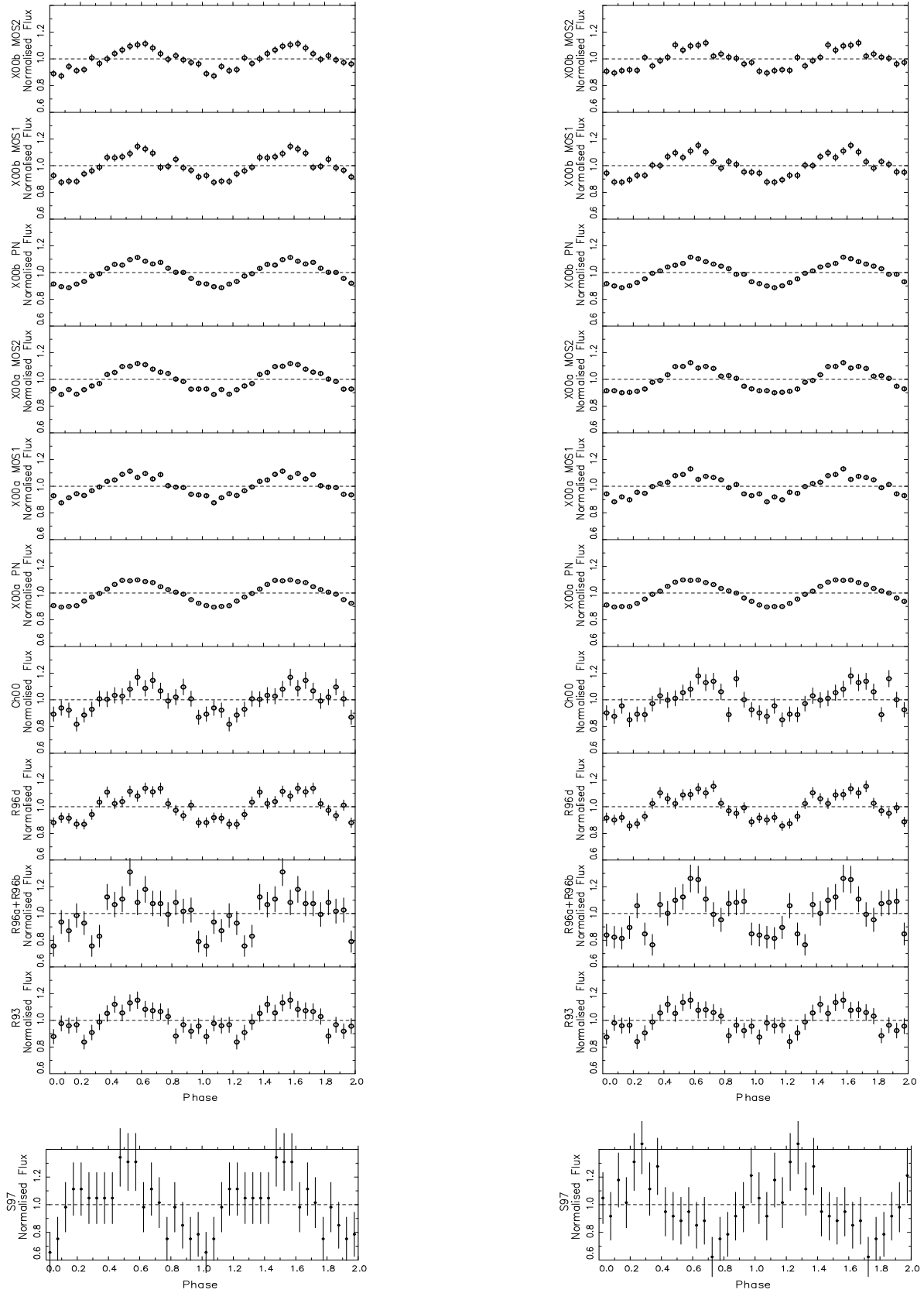

\begin{center}
\setlength{\unitlength}{1cm}
%\fbox{
\begin{picture}(18,22)
\put(-3.5,3){\includegraphics{fig5a.ps}}
\put( 6.0,3){\includegraphics{fig5b.ps}}
\put( 0.8,-3.0){\includegraphics{fig5c.ps}}
\put( 10.3,-3.0){\includegraphics{fig5d.ps}}
\end{picture}
%}
\end{center}
\caption{The datasets folded on $P_0 \dot P$ (1) (left) and (2) (right) in
Table \ref{bestfits}.}
\label{folded}
\end{figure*}

\subsection{Full dataset}
\label{spin:all}

With the further restriction available from the fit to the {\sl Rosat} 1993 and
1996 data subset in Figure \ref{stage2}, a coherent MLP search was made to all
of the data (excluding the uncertain R98 run). The search was made for
$8.391050 \leq P_0 \leq 8.391150$~s and $-3\times10^{-13} \leq \dot P \leq
3\times 10^{-13}$~s/s within the parallel lines $P_0=a\dot P +b$ defined by intercepts $b
=8.391097$~s and 8.391105~s and slope $a= -1.75\times 10^8$ s, 
encompassing the {\sl Rosat} 1993/1996 contours. Care was
taken to ensure
adequate sampling of the $(P_0 \, ,\dot P)$ plane. This identified two
pairs of
values, $(P_0 \, ,\dot P)$ which cannot be discriminated between on
statistical
grounds. These are $(P_0 \, ,\dot P) = 8.39109273$~s, $5.409\times
10^{-14}$~s/s and $
8.39109148$~s, $3.749\times 10^{-14} $~s/s (Table
\ref{bestfits}). All other periods can be excluded at the 95.4\% level,
but two
other possibilites cannot be excluded at the 99\% level: these are also given
in Table \ref{bestfits} for completeness. The contours for the two best fits
are given in Figure \ref{stage3}. 

\begin{table}
\begin{center}
\begin{tabular}{llll}
\hline
Label & $P_0$   & $\dot P$       & $\Delta \chi^2$ \\
      & (s)   &  (s/s)     &               \\
\hline
(1) & $8.39109273$ & $5.409\times10^{-14}$ &    \\
(2) & $8.39109148$ & $3.749\times10^{-14}$ & 1.3  \\
(3) & $8.39108624$ & $6.902\times10^{-14}$ & 6.2  \\
(4) & $8.39108748$ & $8.564\times10^{-14}$ & 6.3  \\
\hline
\end{tabular}
\caption{The four best fit $(P_0 \, ,\dot P)$ values. Refer to Figure
\ref{stage3}
for the confidence interval contours. $\Delta \chi^2$ is the difference between
the $\chi^2$ of a given solution, and that of solution (1). Fits (3) and (4)
are formally excluded at the 95.4\% level.}
\label{bestfits}
\end{center}
\end{table}

We have folded the data on both $P_0 \dot P$ solutions (1) and (2) in
Table \ref{bestfits} to check that the relative phasing of all individual
runs is
correct. We show the folded data on these periods in Figure \ref{folded}. The
data from runs R96a and R96b have been combined to increase the signal-to-noise
ratio, and we have not shown the data from R98 because of the uncertainty in
the time reference for this run (section 2).

Now performing the {\it a postiori} check with the {\sl BeppoSax} data, we find
that these data phase correctly with the main datasets for $P_0 \dot P$ (1),
but not (2). This is shown in the lowest panel of Figure \ref{folded} in each
case. This allows us to select $P_0 \dot P$ (1) as the most likely timing
parameters for RX J0720.4--3125. In any case, both acceptable fits to the data
have $3\times10^{-14} < \dot P < 6\times 10^{-14}$.  For the purposes of our
further discussion, the difference between the $\dot P$ in (1) and (2) is not
important.

\section{Discussion}

The refined value of $\dot P$ reported in this paper is consistent with, but 
two orders 
of magnitude lower than the extrema of the range reported by Haberl et al.
(1997). The first implication is that RXJ0720.4-3125 is unlikely to
be spinning down under propeller torques. If the propeller 
is due to the torque exerted by the interstellar medium, it would be (e.g.
Colpi et al., 1998)
\begin{equation}
\dot P_{prop} \approx 10^{-8} n^{9/13} v_{10}^{-27/13} B_{12}^{8/13}
P^{21/13} \quad {\rm {s \over yr}}   \, ,
\label{prop}
\end{equation}   
where $n$ is the external density in cm$^{-3}$ ($n
\sim 1$ for the interstellar medium), $B_{12}= B
/(10^{12}~{\rm G})$ and
$v_{10}$
is the star's velocity normalized to 10~km/s. By using the values of
$P$ and $\dot P$ of RXJ0720.4-3125, we obtain $B_{12} \approx 16 n^{-9/8}
v_{10}^{27/8}$. Propeller spin-down dominates over
dipolar losses if $B_{12} <  25 {\sqrt n} v_{10}^{-3/2}$ and this,
combined with the previous equation, constrains the star's velocity to
extremely small values: $v_{10} \la n^{1/3}$. Moreover, this scenario 
makes the matching of thermal and dynamical time scales more difficult. On
the other hand, the fact that the propeller contribution is negligeable
supports larger values of the star's velocity.  
                       
In the ``unified'' model suggested by Alpar (2001), the observed
spin-down is associated with propeller effect by a fossil disk. 
In this case, under the assumption that the X-ray luminosity of the source
is supplied by energy dissipation (frictional torque), upper and
lower bounds on $ \dot \Omega$ can be derived by eq.(2) of Alpar
(2001). For RXJ0720.4-3125 the observed luminosity is $L\approx 2 \times
10^{31} (d/100 \, {\rm pc})^2$ erg~cm$^{-2}$~s$^{-1}$, where
$d_{100} = d/100 \, {\rm pc}$ and $d$ is the distance (Haberl et
al. 1997). This gives:
\begin{equation}
2 \times 10^{-12} d_{100}^2  \mincir \dot \Omega
\mincir
2 \times
10^{-10} d_{100}^2  \   { {\rm rad } \over {\rm s}^2} \, ,
\end{equation}
that in turn translates in $\dot P = (P^2 / 2 \pi) \dot \Omega$
\begin{equation}
2 \times 10^{-11} d_{100}^2  \mincir \dot P \mincir 2 \times
10^{-9} d_{100}^2  \   { {\rm s } \over {\rm s} } \, .
\end{equation}

While this scenario is still consistent with the spin-down
measured for RBS 1223 (Hambarayan et al., 2002), the value of
$\dot P$ reported here for RXJ0720.4-3125 is well below this range. Thus, for
RXJ0720.4-3125, the associated energy dissipation cannot,
alone, account for the source luminosity (unless the source is 
at $\sim 5$~pc, which is unrealistic). That also make less plausible 
an interpretation of the hardness ratio profile in terms of a spatially
variable absorbing matter, co-rotating in the magnetosphere (Cropper et
al. 2001). The observed behaviour is more probably explained by the
angle-dependent properties of the emitted radiation. 

On the other hand, the slow spin-down rate of RXJ0720.4-3125 is still
considerable. The other plausible mechanism which may account for such
large and stable
values of $\dot P$ is rotational loss by emission of magnetic radiation.
For a dipolar magnetic field,
\begin{equation}
\dot P \approx 10^{-15} B_{12}^2/P 
\   { {\rm s } \over {\rm s} } \, ,
\end{equation}
which gives a present value of the magnetic field of $B = 2.13
\times 10^{13}$~G \footnote{Note that a twisted magnetosphere may lead
to a reduction up to an order of magnitude in the inferred polar value of
the magnetic field. The considerations about the age, however, remain
unchanged (Thompson, Lyutikov \&
Kulkarni, 2001).}\footnote{Here and in the following we
specify the discussion to solution (1) of Table \ref{bestfits}}. Despite
this
value being below those typically
quoted for magnetars ($B \approx 10^{14}-10^{15}$~G), it is still extreme
and close to the critical value $B_Q =4.41 \times 10^{13}$~G at
which quantizing effects start to be important in shaping the
atmospheric emission. Furthermore, a scenario in which this source is
powered by accretion from the interstellar medium must be ruled out: for
the present values of $P$ and $B$ the corotating magnetosphere will
prevent the incoming material to penetrate below the Alfven radius. 

The corresponding spin-down age is
\begin{equation} 
t_{sd} =  \dot P / \left ( 2P \right
) \sim 2.48 \times 10^6  \, {\rm yr}, 
\end{equation} 
which, given the numerous uncertainties, is marginally compatible with
that inferred by the cooling curves: a few $10^5$ yrs for a surface
temperature of $\sim 80$~eV (see e.g. Kaminker, Haensel \& Yakovlev, 2001,
Kaminker, Yakovlev \& Gnedin, 2001; Schaab et al. 1997, 1999). The
discrepancy is
less significant if we notice that: a) the cooling curves are strongly
dependent on the neutron star mass (Kaminker, Haensel \& Yakovlev, 2001),
and b) what we are probably observing in X-rays is a region of limited
size which
is kept hotter than the average star surface, as inferred by the analysis
of the pulse-shape (Cropper et al., 2001). 

On the other hand, $t_{sd}$ is representative of the true age of the
source only in the case in which the magnetic field remained almost
constant during the star evolution. The same condition applies for the
validity of the cooling curves mentioned above, which do not include the 
extra input of energy released in the neutron star in case of field
decay. It is therefore of fundamental importance to address the field
evolution: the strength of $B$ and its variation during the neutron star
history determine, in fact, the actual luminosity, the lifetime and even
the nature of the energy loss from the star. A related question is where
the field is anchored: in the core, penetrating the whole star or
confined in the crust. There are three
meachanisms which are typically proposed for inducing field-decay: 
ambipolar diffusion in the solenoidal or irrotational mode and Hall
cascade (Goldreich \& Reisenegger 1992, Heyl \& Kulkarni 1998, Colpi,
Geppert \& Page 2000). In reality all the three processes co-exist 
with different timescales, and each of them may dominate the
evolution depending on $B$, $L$ at any given time. In absence of
more detailed
computations, Heyl \& Kulkarni (1998) and Colpi, Geppert \& Page (2000)
tentatively isolated the three processes and proposed some simple,
phenomenological rules to mimic the evolution in the three regimes. We
stress that these descriptions are far from being
comprehensive of all the effects that can modify substantially the
results: they are used in a first
approximation and have the main advantage of having a simple analytical
form. 
By using the expressions of Colpi et al. (2000), 
we have estimated the source age and the value of the magnetic
field at the birth of the neutron star, $B_0$. Results are shown in Table
\ref{decay}. In all cases the
star is assumed to be born with a period of 1 ms: results are not
strongly dependent on this exact value, provided it is far less
than the present period. 

\begin{table}
\begin{center}
\begin{tabular}{lll}
\hline
B-Decay Mechanism & $B_0$ & age \\
      &  $10^{13}$ G &  (years) \\
\hline
Hall Cascade & 119.2 & $4.5 \times 10^4$\\
Ambipolar diffusion, irrotational mode & 1.9 & $3.3 \times 10^6$\\
Ambipolar diffusion, solenoidal mode & 4.1 & $1.6 \times 10^6$\\
\hline
\end{tabular}
\caption{Predicted source age and primordial field for three different
mechanisms of decay, simulated as in Colpi et al. (2000). The present
values of $P$ and $\dot P$ are those of solution (1) in table
\ref{bestfits}. In all cases, the source is assumed to be born
with $P = 1
$~ms.}
\label{decay}
\end{center}
\end{table}

As we can see, allowing for a mechanism involving
very fast decay, such as the Hall cascade, we find that the source is now
$\sim 4 \times 10^4$~yr old, and is born with a superstrong field $B_0
\sim 10^{15}$~G. Such a young age is only marginally compatible with the
absence of a remnant and, more important, is not compatible with the
low observed X-ray luminosity (as we compared using not only the
standard cooling curves mentioned above, but also some cooling curves
kindly provided by Geppert \& Colpi, private communication; these latter
are
computed allowing for the extra-heating due to $B$-decay from Hall
cascade and predict larger luminosities than the former, making
the discrepancy even higher). Underluminous models have been recently
presented by Kaminker, Haensel \& Yakovlev (2001), who accounted for the
enhanced
neutrino cooling in presence of strong neutron superfluidity. These
solutions may match an age of $\sim 10^4$~yrs for RXJ0720.4-3125, but, as
discussed by the same authors, they must probably be rejected since they
fail in
the comparison with observational data of a sample of other neutron stars. 

On the other hand, both mechanisms involving
ambipolar diffusion predict a magnetic field quite stable over the
source lifetime and close to the actual value. Accordingly, the
predicted age is $\sim 10^6$ years in all cases, close to $t_{sd}$. 
Below $\sim 10^{14}$~G the cooling curves are not significantly
influenced by decay through ambipolar diffusion (Heil \& Kulkarni, 1998),
thus, as in the case of constant $B$ discussed above, the scenario is 
compatible with the observed luminosity. The larger age is also compatible
with the absence of a remnant. 

If our conclusions are valid, the connection between dim INS and AXPs is 
not so obvious. RXJ0720.4-3125 has a strong, but not superstrong, field
which is compatible with those of the canonical radio-pulsars which have
passed the death line. On the other hand, having excluded accretion,
what mechanism causes an X-ray emission concentrated in a fraction of
$\sim 60$\% of the star surface remains a mystery, as well as the related
question
about the validity of using the observed blackbody temperature to locate
the source in the cooling diagram. The variation of the surface
cooling temperature with the latitude predicted so far for strong fields
is not large enough and more complicated explanations are required. 

\section{Acknowledgements}

We are grateful to Mat Page for his advice on the maximum likelihood methods,
and to Darragh O'Donoghue for pointing us to the Raleigh transform, and for the
use of his Eagle Fourier transform code used in the preliminary stages of this
work. We are grateful to Monica Colpi for lots of useful discussion, to
Ulrich Geppert for providing the cooling curves and to Sandro Mereghetti
for reading the manuscript.  

\begin{figure*}
\begin{center}
\setlength{\unitlength}{1cm}
\begin{picture}(18,10)
\put(-4,-3.8){\includegraphics{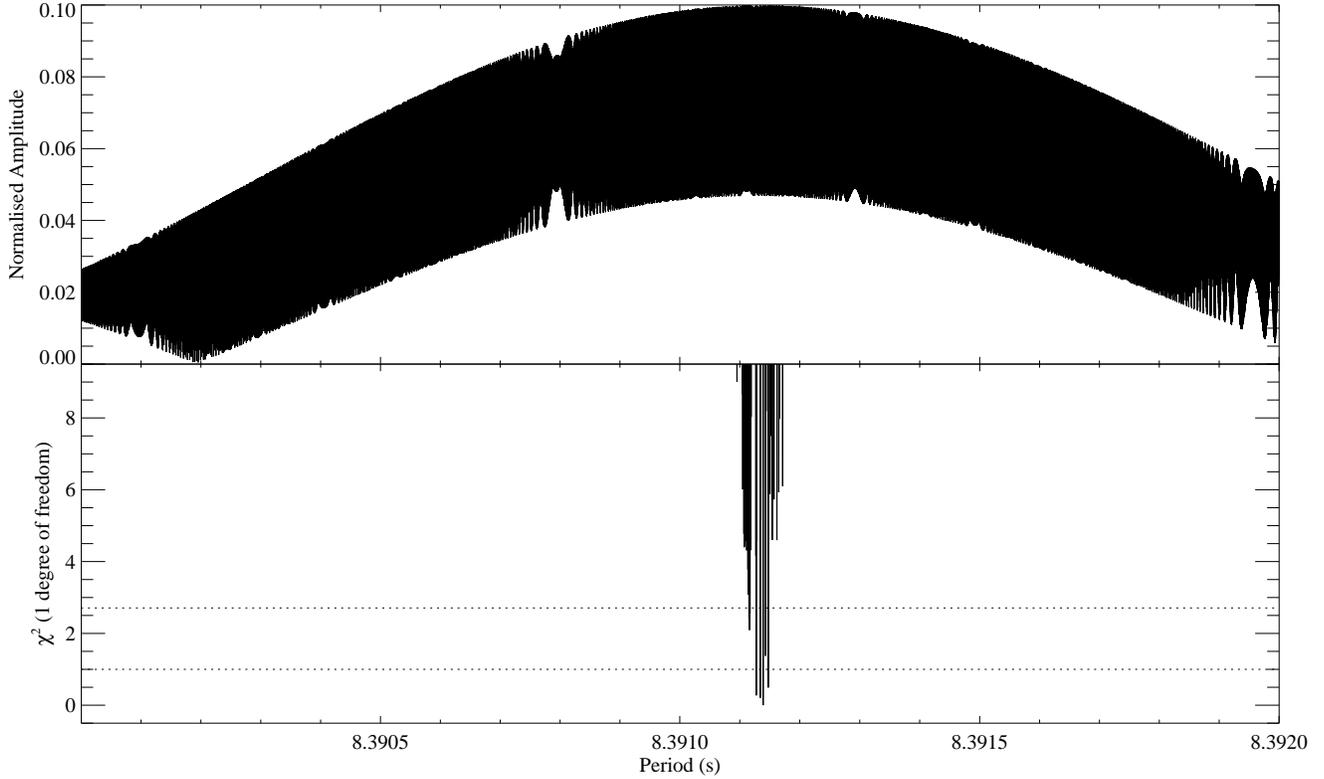}}
\end{picture}
\end{center}
\caption{The discrete Fourier transform (top) and maximum likelihood
periodogramme (bottom) for the X00a PN and X00b PN data over a restricted
period range around the 8.391 s period in RXJ0720.4-3125. The 68\% and 90\%
confidence levels are indicated at $\chi ^2 = 1.0$ and 2.71 for one
degree of freedom.}
\label{fig:appendix}
\end{figure*}

\appendix
\section{Details of the Maximum Likelihood Periodogramme}

The $C$-statistic is derived from the maximum likelihood ratio and is
\begin{equation}
C = 2(E - \sum_{i=1}^{N}n_i \ln I_i)
\end{equation}
(Cash 1979), where $E$ is the sum of the
expected number of counts according to the model distribution, $n_i$ is the
observed number of counts in an interval and $I_i$ is the
expected number of counts at the time of
event $t_i$ according to the model distribution. $N$ is the total number of
counts. For event list data, the
sampling duration tends to zero and $n_i = 1$.
The $\Delta C$-statistic is
\begin{equation}
\Delta C = (C_{min})_{p-q}^T - (C_{min})_p
\end{equation}
where $p$ is the total number of parameters, and $q$ is the number of free
parameters. This is distributed as $\chi^2$ with $q$ degrees of freedom (Cash
1979). $(C_{min})_p$ is the minimum $C$-statistic when all parameters are
varied -- the minimum over all points in the search grid in the free parameter
space. $(C_{min})_{p-q}^T$ is the minimum $C$-statistic at any particular point
in the search grid. Substituting,
\begin{equation}
\Delta C = 2\left[\left(\sum_{i=1}^{N}n_i \ln I_i\right)_{best} -
                  \left(\sum_{i=1}^{N}n_i \ln I_i\right)_{particular}
            \right].
\end{equation}

The expected number of counts $I_i$ in equation (A1) is calculated from the model
distribution. In our case of a pure sinusoidal variation, because the expected
arrival of an event is directly proportional to the model prediction for time
$t_i$, we have
\begin{equation}
I_i = a_0(1 + A\cos(2\pi t_i/P + \theta_0)
\end{equation}
where $A$ is the normalised amplitude, $P$ is the period and $\theta_0$ is the
phase. The scaling of this model $a_0$ is unimportant, as constant factors are
eliminated when substituting in equation (A3). This model has parameters
$(A,P,\theta_0)$ so $p=3$. The periodogramme will scan in $P$, with best fit
values for $A$ and $\theta_0$, so $q=1$. In the case where we allow a period
change,
\begin{equation}
P = P_0 + \dot P t_i
\end{equation}
where $P_0$ is the period at $t_i = 0$ and $\dot P$ is the period change, the
periodogramme will consist of a scan in $P_0$ and $\dot P$. In this case $p=4$
and $q=2$.

In the general case of a maximum likelihood statistic, an optimisation search
is required to obtain the best fit for the $p-q$ fitted parameters. Here, in
the case of a sinusoid, these can be derived directly from the Rayleigh
transform, as pointed out by Bai (1992),
\begin{equation}
z = \frac{1}{N}\left[\left(\sum_{i=1}^{N} n_i \cos{2\pi t_i/P}\right)^2 +
                     \left(\sum_{i=1}^{N} n_i \sin{2\pi t_i/P}\right)^2\right]
\end{equation}
where $z$ is the Rayleigh power, and again, $n_i = 1$ for event list data.
The amplitude is then
\begin{equation}
A = 2\sqrt{z/N}
\end{equation}
and the phase is
\begin{equation}
\theta_0 = \arctan{
          \left[\left( - \sum_{i=1}^{N} n_i \sin{2\pi t_i/P}\right) /
                \left(   \sum_{i=1}^{N} n_i \cos{2\pi t_i/P}\right)\right]
                   }.
\end{equation}
This is equivalent numerically to the discrete Fourier transform (DFT) (Deeming
(1975), as can be ascertained by reference to Kurtz (1985), equation
(A2).

The procedure is then for each $P$ (which may be calculated from equation (A5)
in a $P_0 \, , \dot P$ search) to calculate the amplitude and phase
through
equations (A7) and (A8) (which amounts to calculating the Rayleigh or discrete
Fourier transform), then calculate $\Delta C$ through equations (A4) and
(A3). As the $P$ grid is searched, the maximum $\sum n_i \ln I_i$ is
stored: $\Delta C$ is then simply twice the difference between the calculated
value at each period, and this recorded maximum. This difference is distributed
as $\chi^2$ for one or two degrees of freedom, depending on whether $P$ is used
directly, or calculated in a $(P_0 \, ,\dot P)$ search through equation
(A5).

An illustration of the power of the MLP is given in
Figure~\ref{fig:appendix}. Here we have taken the X00a PN and X00b PN data, and
computed a discrete Fourier transform in the narrow period range 8.390 to 8.392
s (a much expanded scale by comparison with Figure~\ref{stage1}), assuming no
$\dot P$. This (upper plot) shows a multiplicity of aliases created by the long
gap in the data, superimposed on the broad peak corresponding to the duration
of the longest observation (X00a). These are barely resolved in the plot. The
lower plot is the MLP with the 68\% and 90\% confidence levels indicated. This
eliminates most of the aliases, as the narrowness of the distribution is
related to the precision with which the best fit period can be identified
within the broad peak in the upper plot. In the case of the high
signal-to-noise ratio X00a and X00b data, this is significantly narrower than
the width of the peak, which is set by the window function (Scargle 1982).

\end{document}